\documentclass[runningheads]{cl2emult}
\usepackage{makeidx}  
\usepackage{graphicx} 
\usepackage{subeqnar} 
\usepackage{multicol} 
\usepackage{cropmark} 
\usepackage{eso}      
\makeindex            

\begin{document}
\title*{Mining Gamma-Ray Burst Data}
\toctitle{Mining Gamma-Ray Burst Data}
%
%
\titlerunning{Mining Gamma-Ray Burst Data}
%
\author{Jon Hakkila\inst{1}\inst{2}
\and Richard J. Roiger\inst{2}
\and David J. Haglin\inst{2}
\and Robert S. Mallozzi\inst{3}
\and Geoffrey N. Pendleton\inst{3}
\and Charles A. Meegan\inst{4}}
\authorrunning{Jon Hakkila et al.}
%
%
\institute{College of Charleston, SC 29424, USA
\and Minnesota State University, Mankato, MN 56001 USA
\and University of Alabama in Huntsville, AL 35899, USA
\and NASA/MSFC, Huntsville, AL 35812, USA}

\maketitle              

\begin{abstract}
Gamma-ray bursts provide what is probably one of the messiest of all astrophysical data sets. Burst class properties are indistinct, as
overlapping characteristics of individual bursts are convolved with effects of instrumental and sampling biases. Despite these complexities,
data mining techniques have allowed new insights to be made about gamma-ray burst data. We demonstrate how data mining techniques have
simultaneously allowed us to learn about gamma-ray burst detectors and data collection, cosmological effects in burst data, and properties of
burst subclasses. We discuss the exciting future of this field, and the web-based tool we are developing (with support from the NASA AISR
Program). We invite others to join us in AI-guided gamma-ray burst classification (http://grb.mnsu.edu/grb/).
\end{abstract}

\section{Introduction}
Understanding the physics of a class of astronomical objects depends on identifying intrinsic behaviors. When two or more subclasses are present, each subclass is defined in terms of its own intrinsic behaviors. The process of identifying behavioral characteristics is difficult when the objects' observed characteristics (or {\em attributes}) overlap. Such is the case for cosmic gamma-ray bursts (GRBs), which have a large spread in observed attribute values. Some GRB attribute dispersion is intrinsic, some is caused by measurement error, some is due to systematic ({\em e.g.} instrumental and sampling) biases, and some is caused by the presence of subclasses.

GRB subclass behaviors are difficult to delineate from other causes of attribute dispersion. Two GRB subclasses have been known to exist for some time \cite{C74} \cite{K93}, but it has been difficult to assign individual GRBs to a class because of attribute overlap. Class assignment has been complicated even more by the statistical clustering identification of a third GRB subclass \cite{M98}; properties of this third subclass overlap those of the other two.

\begin{figure}[htbp]
\begin{center}\small
\unitlength=0.66mm
\linethickness{0.4pt}
\begin{picture}(245,80)
\put(5,20){\dashbox{1}(45,50)[ct]
       {\shortstack{\\ \\ Data Warehousing}}}
\put(12.5,30){\framebox(30,30)[cc]
       {\shortstack{Pre- \\ Processed \\ GRB Data}}}
\put(42.5,42){\line(1,0){17}}
\put(42.5,48){\line(1,0){17}}
\put(62.5,45){\line(-1,+1){6}}
\put(62.5,45){\line(-1,-1){6}}
\put(55,20){\dashbox{1}(45,50)[ct]
       {\shortstack{\\ \\ Data Mining}}}
\put(62.5,30){\framebox(30,30)[cc]
       {\shortstack{Classifier \\ Shell}}}
\put(93,42){\line(1,0){8.5}}
\put(93,48){\line(1,0){8.5}}
\put(104.5,45){\line(-1,+1){6}}
\put(104.5,45){\line(-1,-1){6}}
\put(105,30){\framebox(30,30)[cc]
       {\shortstack{Scientific \\ and \\ Logical \\ Assessment}}}
\put(135.5,42){\line(1,0){5.5}}
\put(135.5,48){\line(1,0){5.5}}
\put(144.5,45){\line(-1,+1){6}}
\put(144.5,45){\line(-1,-1){6}}

\put(120,30){\line(0,-1){15}}
\put(49.5,14){Iteration}
\put(120,15){\line(-1,0){49}}
\put(47.5,15){\line(-1,0){20}}
\put(27.5,15){\vector(0,+1){15}}

\put(145,30){\framebox(30,30)[cc]
       {\shortstack{GRB \\ Subclasses}}}
\put(0,10){\dashbox{0.5}(140,70)[ct]
       {\shortstack{\\ \\ Gamma-Ray Burst Classification Tool}}}

\end{picture}
\end{center}
\vspace*{-10mm}
\caption{Gamma-Ray Burst Classification Process}\label{fig:fig0}
\end{figure}
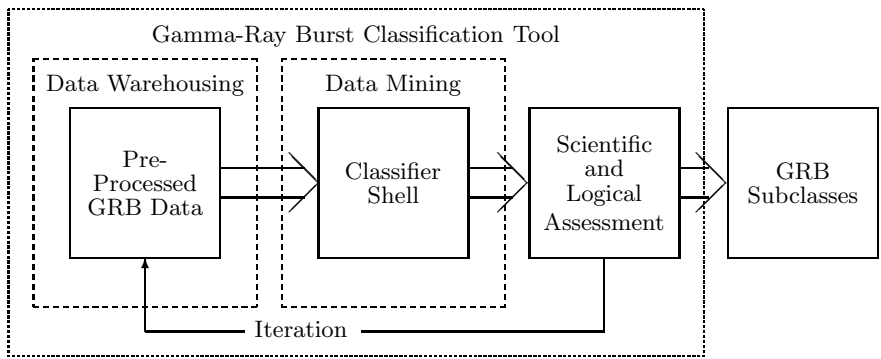

GRB classification can be aided by Knowledge Discovery in Databases (KDD) \cite{D00}. The approach uses pattern recognition algorithms from the Artificial Intelligence (AI) branch of computer science to find behaviors indicative of subclasses. KDD offers a methodology by which meaningful information can be extracted from large volumes of data. The KDD process (Figure \ref{fig:fig0}) is composed of data pre-processing and storage (data warehousing), data mining (clustering software), and scientific/logical assessment. Statistical and systematic effects (e.g. instrumentation and sampling biases) can be identified and even removed in the assessment step.

AI classifiers are typically {\em supervised} or {\em unsupervised}. Supervised classifiers require training {\em instances} (data elements) in order to develop classification rules for unknown instances. Unsupervised classifiers try to subclassify a data set by searching for clusters in multidimensional attribute spaces.

We are developing a web-based tool \cite{Hag00} for the classification of GRB data (http://grb.mnsu.edu/grb/). The tool contains a preprocessed GRB database, AI classifiers, and data visualization software. In this manuscript we describe some of our initial scientific results concerning GRB data mining with this tool. Additional results have been published elsewhere \cite{H00}, \cite{R00}, \cite{Hak00}.

\section{Support for the Existence of Three GRB Subclasses}

Statistical clustering analysis \cite{M98} has revealed the presence of a third GRB subclass for BATSE 3B data \cite{M96}. Three major attributes delineate the three classes; S23 fluence (time-integrated flux in the 50 to 300 keV range), T90 duration (time interval during which 90\% of the burst's emission is received), and HR321 hardness ratio (the fluence in the 100 to 300 keV band divided by the fluence in the 25 to 100 keV band). The properties of the three subclasses are demonstrated in Table \ref{table1}.

\begin{table}
\centering
\caption{Statistical clustering classes, from 3B GRBs.}
\renewcommand{\arraystretch}{1.4}
\setlength\tabcolsep{5pt}
\begin{tabular}{lccc}
\hline\noalign{\smallskip}
Attributes& Class 1 (Long)& Class 2 (Short)& Class 3 (Intermediate) \\
\noalign{\smallskip}
\hline
\noalign{\smallskip} 
T90: & long & short & intermediate \\
S23: & bright & faint & intermediate \\
HR321: & intermediate & hard & soft \\
\hline
\end{tabular}
\label{table1}
\end{table}

We examine the viability of these subclasses using the decision tree classifier C4.5 \cite{Q86}. A {\em decision tree} is a supervised classifier that develops rules by sorting through training instances via a series of branching tests. The results of the tests are turned into IF THEN ELSE statements.

We use C4.5 to demonstrate a new data visualization technique we call ``Fuzzy Controlled Learning'' (or FCL). FCL helps users to visualize the attribute space in which subclasses reside, while recognizing that the subclass distributions overlap in this space. FCL is best used when a principal attribute is available that serves as a performance indicator. We assume for this analysis that T90 duration is the principle attribute, since the longest and shortest GRBs have quite different characteristics.

We withhold 50 GRBs from the long and short ends of the BATSE 3B T90 distribution as ``comparison'' GRBs. These GRBs are considered to have attributes (e.g. fluence and hardnesses) most indicative of the long and short subclasses. Initially, 50 long and 50 short GRBs from the remaining data are used as training instances for C4.5. C4.5 produces first a decision tree and then a rule set for classifying these GRBs. The rules are applied to the comparison bursts; from this rule accuracy is determined.
On each subsequent application, training instances are selected farther from the ends of the T90 distribution; rule accuracies are determined for each training set. The accuracies indicate how closely GRBs in that particular region of the attribute space compare to those in the comparison set; a score near 100\% indicates that training set is indistinguishable from the comparison set, while a score near 50\% indicates that C4.5 could only guess at subclass characteristics. 

\begin{figure}
\centering
\includegraphics[width=.6\textwidth]{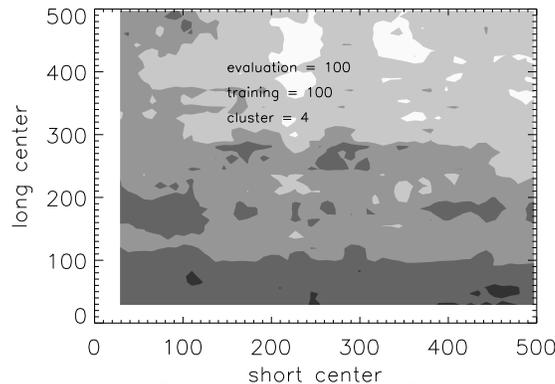}
\caption[]{FCL contour plot. Contours indicate the following agreements between training and comparison data: 90\% (dark), 80\%, 70\%, 50\%, and 30\% (light).}
\label{fig1}
\end{figure}

Figure \ref{fig1} is the FCL contour plot of these rule accuracies. The vertical axis is the distance of the long training cluster center (in units of numbers of GRBs) from the longest GRB, while the horizontal axis is the distance of the short training cluster center from the shortest GRB. The darkest contours (accuracies $\ge 70\%$) near the x-axis indicate that there are several hundred GRBs with long burst characteristics; there are far fewer clearly defined short GRBs near the y-axis. Interestingly, the lightest contours (accuracies $\le 30\%$) occur roughly 200 GRBs from the short end and 350 GRBs from the long end (corresponding to T90s between 4.5 seconds and 16.8 seconds). GRBs in this T90 range have characteristics dissimilar to both long and short bursts; this failure of the two subclass hypothesis to explain the GRB data supports the existence of a third (intermediate) subclass.

\section{Is Each Subclass a Separate Source Population?}

C4.5 is subsequently trained on the three GRB classes \cite{M98} defined from BATSE 3B data \cite{M96}. Several GRBs are found to have peculiar hardness ratios which result from large individual channel fluence errors. 
The GRBs with the largest 10\% relative errors (error divided by 
measurement) are removed, and the remaining 3B 
GRBs are reclassified using C4.5. The resulting rules are used to 
classify 4B Catalog GRBs and thus increase the database size. 

\begin{figure}
\centering
\includegraphics[width=.7\textwidth]{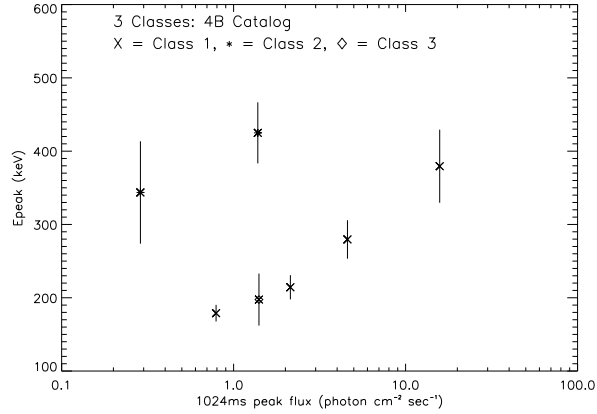}
\caption[]{E$_{\rm peak}$ vs. p1024 for the Three GRB Classes.}
\label{fig2}
\end{figure}

With the larger classification database, the spectral hardness dependence is examined in terms of spectral fitting parameters $\alpha$, $\beta$, and E$_{\rm peak}$ \cite{B93}.
Using only these three attributes, C4.5 accurately classifies most of the 4B GRBs. The resulting rules separate Class 2 from Class 1, but can not delineate Class 3 from Class 1 (85\% of Class 3 GRBs are assigned to Class 1).
Class 3 GRBs are found to have E$_{\rm peak}$ 
values similar to Class 1 bursts of the same peak flux 
(Figure \ref{fig2}). The correlation between E$_{\rm peak}$ and 
peak flux appears due to cosmological redshift \cite{M95}.
Since one of the three defining Class 3 characteristics is a data correlation, we hypothesize that instrumental and/or sampling biases can cause some Class 1 GRBs to take on Class 3 values (e.g. some Class 1 GRBs might appear shorter and fainter than expected).

\begin{figure}
\centering
\includegraphics[width=.7\textwidth]{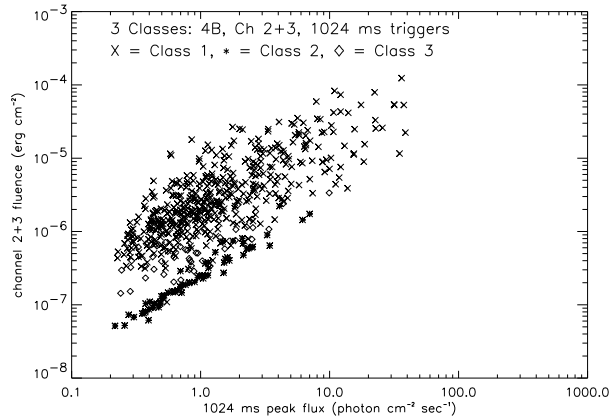}
\caption[]{Fluence vs. p1024 for the Three GRB Classes.}
\label{fig3}
\end{figure}

Figure \ref{fig3} is a plot of fluence vs. 1024 ms peak flux for each 
of the three subclasses, limited to GRBs detected with one homogeneous set of BATSE trigger criteria. There are distinct bounds outside of which no GRBs are found. GRBs with 1024 ms peak fluxes less than 0.2 photons cm$^{-2}$ sec$^{-1}$ are not detected, since this is BATSE's minimum detection threshold. GRBs do not have fluences less than their time-integrated 1024 ms peak fluxes, establishing a lower fluence limit. 

\begin{figure}
\centering
\includegraphics[width=.7\textwidth]{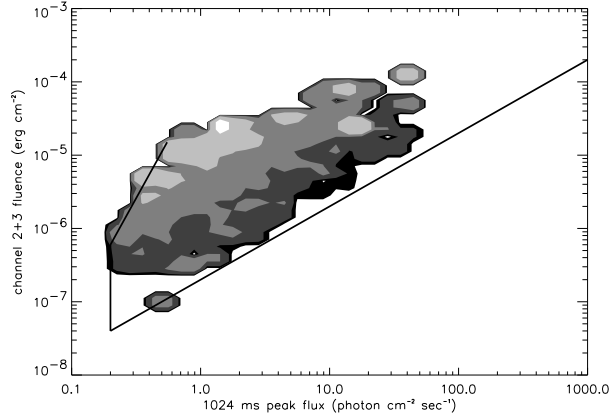}
\caption[]{S23 vs. p1024 for Class 1 GRBs; contours indicate constant log(T90) regions.}
\label{fig4}
\end{figure}

Figure \ref{fig4} overlays $\log$(T90) contours for Class 1 GRBs on the 
fluence vs. 1024 ms peak flux space. The contours demonstrate that 
GRBs can be modeled as a series of pulses, with pulses containing most of 
the fluence and interpulse separations primarily defining the duration. 
Most Class 2 bursts are single-pulsed events as measured on the 1024 
ms timescale.  This helps define the characteristics of the third 
distinct region outside of which no GRBs are found: high fluence, 
faint Class 1 GRBs are missing, whereas low fluence faint, Class 1 
GRBs are present. A bias favoring detection of GRBs with few 
photons over those with many photons seems unlikely, so we suspect a bias 
that removes Class 1 fluence relative to peak flux.

We have dimmed five temporally different GRBs through ten peak fluxes in order to study their measured properties as they fade into background. Each dimmed GRB's time history is Poisson ``noisified,'' then the peak flux and fluence are re-measured. 

\begin{figure}
\centering
\includegraphics[width=.7\textwidth]{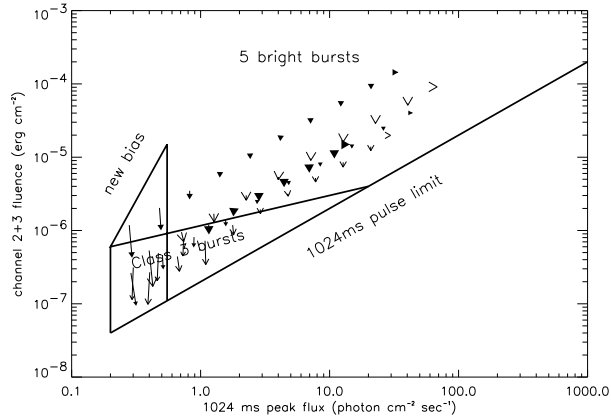}
\caption[]{Fluences and peak fluxes of five decremented and noisified Class 1 GRBs (durations taken from identifiable pulses).}
\label{fig5}
\end{figure}

The time interval bounding the fluence measurement (the {\it fluence duration} 
\cite{Hak00}) appears to strongly influence the amount of fluence measured. 
If the same fluence duration interval is used for undimmed and dimmed 
measurements, then the fluence-to-peak flux ratio does not change as a GRB 
is dimmed. If, however, the fluence duration interval is shortened to 
account for faint pulses disappearing into the background 
and becoming unrecognizable, then the fluence-to-peak flux ratio 
decreases as the burst is dimmed (see Figure \ref{fig5}). This bias is
stronger near trigger threshold. 

\begin{figure}
\centering
\includegraphics[width=.6\textwidth]{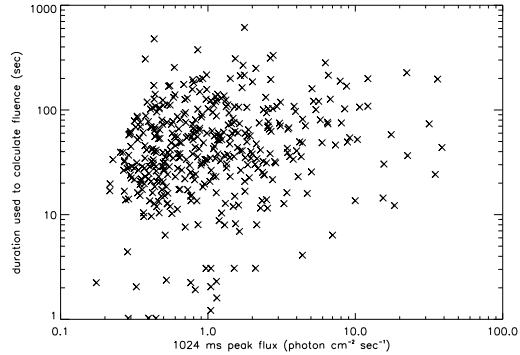}
\caption[]{Fluence Durations of Class 1 GRBs.}
\label{fig6}
\end{figure}

Fluence durations taken from BATSE Catalogs provide supportive evidence 
for this mechanism (see Figure \ref{fig6}). Fluence durations of faint 
Class 1 GRBs are shorter than those of bright Class 1 GRBs \cite{Hak00}.

\section{Conclusions}

We have demonstrated that data mining techniques can aid the interpretation of scientific data, even with complex and ambiguous GRB data.  Data mining demonstrates that some Class 1 (Long) GRBs can develop 
Class 3 (Intermediate) characteristics via a combination of the 
hardness intensity relation and the fluence duration bias. Class 3 
(Intermediate) GRBs do not appear to represent a separate 
source population, although they cluster in the duration, fluence, 
hardness, attribute space. Class 2 (Short) GRBs do appear to represent 
a separate source population.

\clearpage
\addcontentsline{toc}{section}{Index}
\flushbottom
\printindex


\begin{thebibliography}{7}
%
\addcontentsline{toc}{section}{References}

\bibitem{B93} Band, D. L. et al. (1993)
Ap. J. {\bf 413}, 281--292

\bibitem{C74} Cline, T. L., Desai, U. D. (1974) 
Proc. 9th ESLAB Symp. ESRO, Noordwijk, p. 37--45

\bibitem{D00} Djorgovski, G. J. (2000), this conference

\bibitem{Hag00} Haglin, D. J. et al. (2000) 
In Gamma-Ray Bursts, ed. M. Kippen, R. S. Mallozzi, \& G. J. Fishman (AIP: New York) 877--881

\bibitem{H00} Hakkila J. et al. (2000)
Ap. J. {\bf 538}, 165--180

\bibitem{Hak00} Hakkila J. et al. (2000)
In Gamma-Ray Bursts, ed. M. Kippen, R. S. Mallozzi, \& G. J. Fishman (AIP: New York) 48--52

\bibitem{K93} Kouveliotou, C. et al. (1993)
Ap. J. {\bf 413}, L101--L104

\bibitem{M95} Mallozzi, R. S. et al. (1995)
Ap. J. {\bf 454}, 597--603

\bibitem{M96} Meegan, C. A. et al. (1996)
Ap. J. S. {\bf 106}, 65--110

\bibitem{M98} Mukherjee, S. et al. (1998)
Ap. J. {\bf 508}, 314--327

\bibitem{P99} Paciesas, W. S. et al. (1999)
Ap. J. S. {\bf 122}, 465--495

\bibitem{Q86} Quinlan, J. R. (1986)
Machine 
Learning {\bf 1}, 81--106

\bibitem{R00} Roiger, R. J. et al. (2000)
In Gamma-Ray Bursts, ed. M. Kippen, R. S. Mallozzi, \& G. J. Fishman (AIP: New York) 38--42

\end{thebibliography}
\end{document}